\documentstyle[aps,prb,epsf,eqsecnum,floats]{revtex}
\begin{document}
\twocolumn[\hsize\textwidth\columnwidth\hsize\csname@twocolumnfalse%
\endcsname

\title{Quantum Phases of the Shastry-Sutherland Antiferromagnet}
\author{C. H. Chung and J. B. Marston}
\address{Department of Physics, Brown University, Providence, RI
02912-1843}
\author{Subir Sachdev}
\address{Department of Physics, Harvard University, Cambridge MA
02138\\ and Department of Physics, Yale University, P.O. Box
208120, New Haven, CT 06520-8120}
\date{February 12, 2001}

\maketitle

\begin{abstract}
We study possible paramagnetic phases of antiferromagnets on the
Shastry-Sutherland lattice by a gauge-theoretic analysis of
fluctuations in a theory with Sp($2N$) symmetry. In addition to
the familiar dimer phase, we find a confining phase with plaquette
order, and a topologically ordered phase with deconfined $S=1/2$
spinons and helical spin correlations. The deconfined phase is
contiguous to the dimer phase, and in a regime of couplings close
to those found in the insulator SrCu$_2$(BO$_3$)$_2$. We suggest
that a superconductor obtained by doping this insulator with
mobile charge carriers will be an attractive candidate for
observing the anomalous magnetic flux properties associated with
topological order.

\end{abstract}
\pacs{PACS numbers:}

]



\section{Introduction}
\label{intro}

Much interest has recently focused on the magnetic properties of
the insulator SrCu$_2$(BO$_3$)$_2$\cite{Kageyama,Onizuka}. The
low energy spin excitations in this material reside on the $S=1/2$
Cu ions which lie in two-dimensional layers decoupled from each
other. The experiments show a clear indication of an energy gap
towards spin excitations, making this one of the few known
two-dimensional systems with a spin gap. Remarkably, the pattern
of near-neighbor antiferromagnetic exchange couplings between the
Cu ions turns out to be identical to that in a model Hamiltonian
studied many years ago by Shastry and Sutherland\cite{Shastry}.
These authors also showed that a simple decoupled dimer
wavefunction was an exact eigenstate of this Hamiltonian, and
that it was the ground state over a restricted parameter regime.

\begin{figure}
\epsfxsize=2.4in \centerline{\epsffile{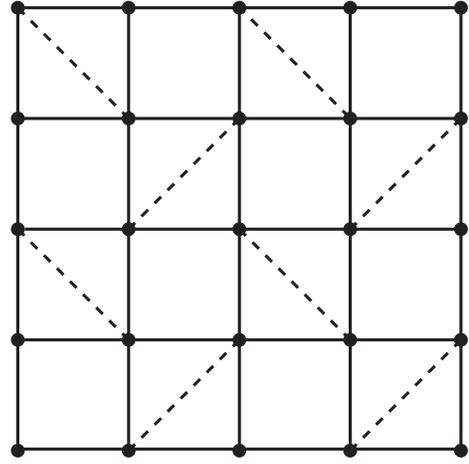}} \vspace{12pt}
\caption{The Shastry-Sutherland lattice. The exchange $J_1$ acts
between sites separated by the horizontal and vertical links,
which the exchange $J_2$ acts across the diagonal links.}
\label{fig1}
\end{figure}

The Shastry-Sutherland antiferromagnet is sketched in
Fig~\ref{fig1}. The Hamiltonian is
\begin{equation}
H = J_1 \sum_{\langle ij \rangle} {\bf S}_i \cdot {\bf S}_j + J_2
\sum_{\mbox{diagonals}} {\bf S}_i \cdot {\bf S}_j
\label{SSH}
\end{equation}
where ${\bf S}_i$ are $S=1/2$ operators on the sites, $i$, of a
square lattice. The exchange $J_1 > 0$ acts along the nearest
neighbor links (shown as full lines in Fig~\ref{fig1}), while $J_2
> 0$ acts on the diagonal links, shown as dashed in lines in
Fig~\ref{fig1}. It was established\cite{Shastry} that a simple
product of singlet pairs on the diagonal links was the ground
state of $H$ for sufficiently large $J_2/J_1$. However, an
understanding of the experiments requires a description of the
excitation spectrum, and also of possible quantum phase
transitions to other states at smaller $J_2/J_1$. These issues
have been addressed in a number of recent theoretical
works\cite{Miyahara,Mila,Weihong,Muller,kk,uhrig,cb,girvin}. Many
of these studies\cite{Miyahara,Weihong,Muller,kk} involve
numerical analyses based upon large-order series expansions
departing from various decoupled cluster states. Quantum Monte
Carlo simulations have, in principle, a smaller bias due to the
choice of an initial state, and can be extended to much large
system sizes; however, simulations of $H$ suffer from a sign
problem, and so such studies have not been possible. An analytic
mean-field approach has also been undertaken by Albrecht and
Mila\cite{Mila}: results were obtained mainly for the magnetically
ordered states, and the various distinct paramagnetic states were
not distinguished.

Quite apart from determining the ground states of the specific
Hamiltonian $H$, it is also of interest to determine the phases of
models which are ``near'' the parameter space of $H$. This is in
the hope that future experiments may succeed in deforming the
insulator SrCu$_2$(BO$_3$)$_2$ by substitutional doping (which can
induce mobile carriers), or by the application of hydrostatic
pressure. Doping the antiferromagnetic insulator La$_2$CuO$_4$ led
to the discovery of high temperature superconductivity: related
phenomena may be expected here, although, as we shall argue later,
the presence of strong frustration in the parent insulator
SrCu$_2$(BO$_3$)$_2$ may lead to profound differences in the
nature of a possible superconducting ground state.

This paper will examine a generalization of $H$ to Sp($2N$)
symmetry (SU(2) $\cong$  Sp(2)) and describe the properties of the
large $N$ limit. Some of the phases obtained in such a large $N$
limit may not actually appear in the phase diagram of the SU(2)
model $H$--- nevertheless, as we have just argued, the phases may
still be of relevance to physical systems whose microscopic
Hamiltonians are near the parameter space of $H$. Such an approach
has been fruitfully applied to a number of other frustrated
quantum antiferromagnets in previous
work\cite{rsprl,rodolfo,rsijmpb,triangle,chung}. The method leads
to an unbiased selection of possible ground states in the large
$N$ limit, both with and without broken spin rotation symmetry.
Moreover, a gauge-theoretic description of the fluctuations about
the mean-field solution allows a systematic and reliable
assessment of the stability of the various ground states, along
with a description of the dynamics of the excitations.

The Sp($2N$) generalization of $H$ is defined by introducing
canonical Bose creation operators $b_{i \alpha}^{\dagger}$ on
every site $i$, with $\alpha = 1 \ldots 2N$ a Sp($2N$) index. The
allowed states in the Hilbert space satisfy the constraint
\begin{equation}
b_{i \alpha}^{\dagger} b_{i}^{\alpha} = 2NS \label{constraint}
\end{equation}
on every site $i$ (we follow the convention of summing over all
repeated Sp($2N$) indices); the right hand side of
(\ref{constraint}) must be a positive integer, the values of $S$
are constrained accordingly---for the physical case, $N=1$, $S$
must take half-integral values, as expected. The Hamiltonian is
\begin{eqnarray}
H &=& - \frac{J_1}{2N} \sum_{\langle ij \rangle} \Bigl( {\cal
J}^{\alpha\beta} b_{i\alpha}^{\dagger} b_{j \beta}^{\dagger}
\Bigr) \Bigl( {\cal J}_{\gamma\delta} b_{i}^{\gamma}
b_{j}^{\delta} \Bigr) \nonumber
\\ &-& \frac{J_2}{2N} \sum_{\mbox{diagonals}} \Bigl( {\cal
J}^{\alpha\beta} b_{i\alpha}^{\dagger} b_{j \beta}^{\dagger}
\Bigr) \Bigl( {\cal J}_{\gamma\delta} b_{i}^{\gamma}
b_{j}^{\delta} \Bigr) \label{hamitonian}
\end{eqnarray}
where ${\cal J}^{\alpha\beta} = {\cal J}_{\alpha\beta} = - {\cal
J}_{\beta \alpha}$ is the generalization of the antisymmetric
$\varepsilon$ tensor of SU(2) ({\em i.e.}) ${\cal J}$ contains
$N$ copies of $\varepsilon$ along its center block diagonal, and
vanishes elsewhere).

The large $N$ analysis of a large class of models, of which $H$ is
a member, was described with some generality in Section II of
Ref.~\onlinecite{triangle}. We will follow the same method here,
and so will dispense with the details of the computation. The
resulting mean-field phase diagram is shown in Fig~\ref{phase} as
a function of $J_2/J_1$ and $1/S$ (in the large $N$ limit, $S$
becomes a continuous real variable). The positions of the various
phase boundaries are not expected to be quantitatively accurate
for the physical $N=1$ case. However, the general topology of the
phase diagram, the nature of the phases and their excitations,
and the critical properties of the quantum phase transitions can
be reliably described using Fig~\ref{phase} as a starting point.

The properties of all the phases in Fig~\ref{phase} will be
discussed in detail in Section~\ref{mft}. Here we highlight our
main new results.

One of the paramagnetic phases has short-range equal-time spin
correlations peaked at the wavevector $(\pi, \pi)$. [We denote
this phase $(\pi, \pi)$ short-range ordered (SRO)
in Fig~\ref{phase}; here we are placing the sites on
the vertices of a regular square lattice as in Fig~\ref{fig1},
and measuring wavevectors in units of $1/(\mbox{nearest neighbor
spacing})$.  In the experimental SrCu$_2$(BO$_3$)$_2$ system, the
positions of the sites is different, and there will be a
corresponding transformation in the wavevector dependence of
observables.]  At the mean field level, this phase is identical to
that found earlier\cite{peierls1,peierls2} on the square lattice
with $J_2=0$. However, we will show here that a difference does
emerge upon consideration of fluctuations. For $J_2=0$, it was
shown\cite{peierls1,peierls2} that Berry phases associated with
hedgehog-instantons led to columnar spin-Peierls order in the
$(\pi, \pi)$ SRO phase. Here we show that a closely related
analysis for the Shastry-Sutherland lattice leads instead to
``plaquette'' order in this phase. Just such a phase was
considered recently by Koga and Kawakami\cite{kk}.

Our other new results are also associated with a paramagnetic
phase. This phase is denoted $(\pi, q)$ SRO in Fig~\ref{phase} and
is obtained by a destroying the long-range magnetic order in a
helically ordered phase.  Equal-time spin correlations show
short-range incommensurate order, and the spin structure factor is
peaked at the incommensurate wavevector $(\pi, q)$ (the value of
$q$ varies continuously as a function of $J_2/J_1$). As in
previous incommensurate SRO phases found on frustrated square
lattice models\cite{rsprl,rsijmpb}, we argue that the excitations
above the ground state are {\em deconfined spinons} which carry
spin $S=1/2$ (for SU(2)). Also as in previous
work\cite{rsprl,rsijmpb}, the quantum phase transition between
this phase and the $(\pi,\pi)$ SRO (plaquette) phase
(Fig~\ref{phase}) is described by theory of a charge 2 Higgs
scalar coupled to a compact $U(1)$ gauge field; the deconfinement
transition is associated with the condensation of the Higgs
field, and the critical properties are those of a $Z_2$ gauge
theory\cite{eduardo,rsprl,rodolfo,rsijmpb,wen,vojta,senthil}. We
will also consider here the transition between the deconfined
phase and the dimer phase: by a somewhat different analysis, we
will show that this transition also reduces to a $Z_2$ gauge
theory description.

\section{Mean field phase diagram}
\label{mft}

As discussed in Section II of Ref.~\onlinecite{triangle}, a key
quantity determining the nature of the phases is a complex,
directed, link field $Q_{ij}=-Q_{ji}$. Operationally, this field
is introduced to decouple the quartic boson interactions in $H$ by
a Hubbard-Stratonovich transformation.  After this decoupling, the
effective action contains the terms
\begin{equation}
{\cal S} = \int d\tau \sum_{i>j} \frac{J_{ij}}{2} \left[ N
|Q_{ij}|^2 - Q_{ij}  {\cal J}_{\alpha\beta} b_{i}^{\alpha}
b_{j}^{\beta} + \mbox{H.c.} \right] + \ldots, \label{action}
\end{equation}
where $\tau$ is imaginary time, $J_{ij} = J_1$ ($J_{ij}=J_2$) on
the horizontal/vertical (diagonal) links, and the ellipses
represent standard terms which impose the canonical boson
commutation relations and the constraint (\ref{constraint}). It is
also clear from the structure of ${\cal S}$ that the average value
of $Q_{ij}$ satisfies
\begin{equation}
\left\langle Q_{ij} \right\rangle = \frac{1}{N} \left\langle
 {\cal J}^{\alpha\beta}
b_{i\alpha}^{\dagger} b_{j \beta}^{\dagger} \right \rangle.
\label{meanQ}
\end{equation}
For larger values of $S$, the dynamics of ${\cal S}$ requires
condensation of the $b_{i}^{\alpha}$ bosons, and hence a non-zero
value of
\begin{equation}
x_i^{\alpha} = \langle b_i^{\alpha} \rangle; \label{defx}
\end{equation}
such phases break the spin rotation symmetry, and have magnetic
long-range order. As described in Ref~\onlinecite{triangle}, we
optimized the ground state energy with respect to variations in
$\langle Q_{ij} \rangle$ and $x_i^{\alpha}$ for different values
of $J_2 / J_1$ and $S$.  The four-site unit cell of the
Shastry-Sutherland lattice, depicted in Fig ~\ref{unit}, has 10
different $Q_{ij}$ fields.  Care must be taken to identify
gauge-equivalent configurations.  We find that each saddle point
may be described by purely real $\langle Q_{ij} \rangle$.
The resulting phase diagram is
shown in Fig~\ref{phase}. We describe the phases in turn in the
following subsections, considering first the magnetically ordered
phases with $x_i^{\alpha} \neq 0$ in Subsection~\ref{magnet},
and then the paramagnetic phases in Subsection~\ref{para}.

\subsection{Magnetically ordered phases}
\label{magnet}

\subsubsection{N\'eel $(\pi, \pi)$ LRO state}
\label{neel} This is the familiar long-range ordered (LRO) state
in which $\langle {\bf
S}_i \rangle$ is collinearly polarized in opposite directions on
two checkerboard sublattices.  It is known to be the ground state
of $H$ for $J_2=0$, $S=1/2$ in the physical $N=1$ limit.
A gauge may be chosen in which the expectation values of link
variables, $\langle Q_{ij} \rangle$, are nonzero and equal on the
horizontal and vertical links, while the expectation values on the
diagonal links are zero.  In the notation of Fig ~\ref{unit} then
$Q_i = P_i$ ($i = 1, 2, 3, 4$) and $R_1 = R_2 = 0$.

\subsubsection{Helical $(\pi, q)$ and $(q, \pi)$ LRO states}
\label{helical} This phase is characterized by non-zero values of
$\langle Q_{ij} \rangle$ on the horizontal, vertical and diagonal
links. A gauge choice sets all the $Q_i$ equal to each other, and
similarly for the $P_i$; in the appendix we present an argument
which shows the values of the $P_i$ and $Q_i$ are also equal to
each other. There are two gauge non-equivalent choices for the
values of $R_{1,2}$: one state has $R_1 = R_2$ and the other $R_1
= -R_2$. The two states are interchanged under a $90^\circ$
rotation, and correspond to spirals ordered in the horizontal or
vertical directions. At large values of the spin, this phase
appears at $J_2
> J_1$, in accord with the classical calculation of Shastry and
Sutherland\cite{Shastry}. Equal-time spin correlations exhibit
long-range incommensurate order, and the spin structure factor
peaks at the incommensurate wavevectors $(\pi, q)$ or $(q, \pi)$
with the value of $q$ varying continuously as a function of
$J_2/J_1$. This state also appears in the studies of
Refs~\onlinecite{Mila,cb}.

\subsection{Paramagnetic phases}
\label{para} In this subsection we discuss the three phases for
which $x_i^{\alpha} = 0$. As a consequence, spin rotation symmetry
is preserved and only spin SRO arises; however there may be
ordering in other singlet order parameters.

\subsubsection{Dimer state}
\label{dimer} This is the exact SU(2) eigenstate of decoupled
singlet pairs found by Shastry and Sutherland\cite{Shastry}. In
the large $N$ limit, this corresponds to a saddle point at which
the $\langle Q_{ij} \rangle$ are non-zero only on the diagonal
links: $R_1 = R_2 \neq 0$ and $Q_i = P_i = 0$.
Note that the $b_i^{\alpha}$ boson are spatially decoupled
at such a saddle point: each $b_i^{\alpha}$ can only hop across a
single diagonal link. This simplifies analysis of fluctuations
about the saddle point in or near the dimer state, as will be
discussed in Section~\ref{z2}. At higher orders in $1/N$, the
$b_i^{\alpha}$ can indeed hop through the entire lattice; we
expect that the lowest lying excitation will be a $S=1$ spin
triplet\cite{Miyahara} (for $SU(2)$), consisting of a confined
pair of the $b_{i}^{\alpha}$ bosons.

\subsubsection{$(\pi, \pi)$ SRO}
\label{ppsro} This state is obtained by quantum-disordering the
N\'eel state of Subsection~\ref{neel}, and
the expectation values of $Q_{ij}$ have the same structure as those
in Subsection~\ref{neel}. As has been discussed in some detail in
Refs~\onlinecite{peierls1,peierls2}, the quantum fluctuations in
this phase are described by a compact $U(1)$ gauge theory. Such a
theory is always confining, and thus the $b_i^{\alpha}$ bosons again
bind to yield a $S=1$ quasiparticle above a spin gap. There is
also an interesting structure in the spin-singlet sector: this is
considered in Subsection~\ref{plaquette} where it is demonstrated
that at finite $N$ this phase has ``plaquette'' order.

\subsubsection{$(\pi, q)$ and $(q, \pi)$ SRO}
\label{pqsro} In this phase $\langle Q_{ij} \rangle$ are non-zero
on the diagonal, horizontal and vertical links, like the helical
$(\pi, q)$ LRO phase of Subsection~\ref{helical}. Again there are
two gauge non-equivalent configurations, corresponding to the
choices $R_1 = R_2$ with $Q_i = Q < P_i = P$ [the $(\pi, q)$
phase], and $R_1 = -R_2$ with $Q_i = Q > P_i = P$ [the $(q, \pi)$
phase]. Thus all of the horizontal $Q_i$ fields acquire the same
expectation value, but unlike in the helical LRO phase, this value
differs slightly from that of the vertical $P_i$ fields; the
difference is only on the order of a part in ten thousand. The
state is a spin-singlet and there is a gap to all spin
excitations. Nevertheless, the symmetry of $90^{\circ}$ rotations
between the vertical and horizontal directions is broken---this
would now be apparent in various spin-singlet observables like the
bond exchange energies or the bond-charge densities. This phase
may therefore be viewed as a spin-singlet ``nematic'' as only
rotational symmetry is broken. The choice of a vertical or
horizontal spatial polarization in the nematic order leads to a
two-fold degeneracy in the ground state. The state also has
``topological''
order\cite{rk,bulbul,rsprl,senthil,misguich,moessner}, and this
would lead to an additional four-fold degeneracy in a torus
geometry. Unlike the commensurate SRO phases, the spinons are
deconfined. We describe the deconfinement transition below in
Section~\ref{z2}. The spinon dispersion has its minima at momentum
$(\pi/2, q/2)$ or $(q/2, \pi/2)$. Although this phase is realized
only for $S < 1/2$ in the large $N$ limit, it seems possible that
in the physical limit $N = 1$ it could extend up to $S=1/2$ for a
narrow range of $J_2/J_1$. Similar behavior was found in a study
of the Sp($2N$) Heisenberg antiferromagnet on the anisotropic
triangular lattice\cite{chung}. It would interesting to search for
this phase using numerical methods.

We conclude this section by briefly comparing our results to other
published calculations.  For $S = 1/2$ we find that the transition
between N\'eel and Helical LRO phases is continuous, occurring at
$J_2/J_1 \approx 1.02$, close to the value of $1.1$ found by
Albrecht and Mila\cite{Mila}, who also report a continuous
transition.  Also in agreement with Albrecht and Mila, we find
that the transition between the Helical LRO and Dimer SRO phase is
first order, but occurs at $J_2/J_1 \approx 2.7$ instead of
$1.65$. Carpentier and Balents \cite{cb} also found a helical LRO
phase, but presented arguments that an intermediate phase may
exist between the helical LRO and dimer phases: our $(\pi, q)$ SRO
state is precisely such a phase. Koga and Kawakami\cite{kk}
employed a series expansion to find, for $S=1/2$, a plaquette
phase which intervenes between the N\'eel and dimer phases.  As
shown below, the $(\pi, \pi)$ SRO phase acquires plaquette order
at finite $N$, but as can be seen in Fig ~\ref{phase}, at large
$N$ this phase only occurs for $S < 1/5$. If finite $N$
fluctuations push the phase boundary up to $S = 1/2$ then the
following sequence of phases would occur as $J_2/(J_1+J_2)$
increases from $0$ to $1$: N\'eel, Plaquette $(\pi, \pi)$ SRO,
$(\pi, q)$ SRO, and finally Dimer SRO.

\section{Plaquette order in the commensurate paramagnet}
\label{plaquette}

This section will discuss the fate of the spin singlet sector
upon including fluctuations about the mean-field in the $(\pi,
\pi)$ SRO state. The results below are a straightforward
generalization of those obtained in
Refs~\onlinecite{peierls1,peierls2} for the square lattice
antiferromagnet. We will only consider the case where $2SN$ is an
odd integer (for the physical $SU(2)$ case, this means that $S$
is half an odd integer); the generalization to other values of
$S$ follows as in earlier work.

In the present large $N$ approach, regular perturbative
corrections order by order in $1/N$ do not qualitatively modify
the nature of the mean-field ground state. However, singular
effects do appear\cite{peierls1,peierls2} upon considering the
consequences of `hedgehog' like instanton tunneling events and
their Berry phases. Such a calculation is technically involved,
and a somewhat more transparent discussion of essentially the
same physics emerges from studying the ``quantum dimer''
model\cite{rk} (see Appendix A of Ref~\onlinecite{peierls2} for a
discussion of the equivalence between the instanton physics of
the large $N$ expansion and dual representations of the quantum
dimer model). Here we shall follow the treatment of
Ref~\onlinecite{cavo}.

The quantum dimer model represents the Hilbert space of low-lying
singlet excitations by assuming that it can be mapped onto states
represented by a near-neighbor singlet bond (`dimers') covering
of the lattice. In the present $(\pi, \pi)$ SRO phase, we need
only take dimers connecting nearest neighbor sites on horizontal
and vertical links. The dimers along the diagonal links are
assumed to occur only rarely in this phase: they can therefore be
integrated out, and serve mainly to modify the effective
Hamiltonian in the space of horizontal and vertical dimers.
Indeed, the most important consequence of this procedure is
apparent from a glance at Fig~\ref{fig1}: the diagonal dimers
divide the plaquettes of the square lattice into two classes,
those with and without diagonal links across them, and we expect
dimer resonance terms around these plaquettes to have distinct
matrix elements (see Fig~\ref{figdimer}). This distinction will
be the only difference from earlier analyses\cite{peierls1,peierls2},
and we will show that it is sufficient
to lead to plaquette order in the $(\pi, \pi)$ SRO phase.

Our results emerge from an analysis of the `height'
representation of the quantum dimer
model\cite{peierls2,fradkiv,zheng,cavo,moessner}. There is a
rigorous, one-to-one mapping between the set of coverings of the
square lattice with nearest-neighbor horizontal and vertical
dimers, and the configurations of an interface of heights, $h_a$,
defined on the sites, $a$, of the dual square lattice (we
identify two interfaces is equivalent if they are related by a
uniform translation  $h_a \rightarrow h_a + p$, where $p$ is any
integer). The values of $h_a$ are restricted to
\begin{equation}
h_a = n_a + \zeta_a \label{heights}
\end{equation}
where $n_a$ is a integer which fluctuates from site to site, and
$\zeta_a$ is a fixed fractional offset which takes the values $0,
1/4, 1/2, 3/4$ on four dual sublattices, $X,Y,Z,W$, as shown in
Fig~\ref{figzeta}. We further restrict the $h_a$ to satisfy $|h_a
- h_b| < 1$ for any pair of nearest-neighbor sites $a,b$. We can
now specify the connection between the height model and the dimer
coverings. Examine the value of $|h_a-h_b|$ for every nearest
neighbor pair, and if $|h_a - h_b |>1/2$, place a dimer on link
shared by the plaquettes of the direct lattice around $a$ and $b$.
It is not difficult to see that a consequence of our choice of
the $\zeta_a$ offsets is that dimers so obtained will form a
close-packed covering of the lattice. Examples of the relationship
between the height values and dimer coverings are shown in
Fig~\ref{figdimer}.

We can now use general symmetry considerations to write down an
effective action for the height degrees of freedom. As is
standard in theories of interface models, we promote discrete
heights $h_a$, in (\ref{heights}), to continuous real variables
$\chi_a$ by the Poisson summation formula, and ``soften'' the
constraints to periodic cosine potentials which have minima at
the values $\chi_a = h_a$ which obey (\ref{heights}). In this
manner we obtain the action
\begin{eqnarray}
{\cal S}_{\chi} &=& \int  d \tau \Biggl[ \frac{K}{2}
\sum_{\langle ab \rangle} (\chi_a
- \chi_b)^2 \nonumber \\
&+& \sum_a \left\{ \frac{K_{\tau}}{2} (\partial_{\tau} \chi_a)^2 -
y_a \cos(2\pi(\chi_a - \zeta_a)) \right\} \Biggr], \label{schi}
\end{eqnarray}
where the sum over $\langle ab \rangle$ extends over nearest
neighbor sites, and $K$ is the stiffness towards spatial
fluctuations of the interface height. The corresponding stiffness
towards time-dependent fluctuations is $K_{\tau}$, and, for
simplicity, we have taken its value $a$ independent. The symmetry
of the lattice requires that the strength of the periodic
potential take two possible values, $y_a = y_1$ or $y_a = y_2$
depending upon whether the plaquette $a$ has a diagonal $J_2$
link across it or not. This is the sole distinction from the
analysis of the square lattice antiferromagnet in
Ref~\onlinecite{peierls2}, which had $y_1 = y_2$.

The fundamental property of interface models in 2+1 dimensions,
like ${\cal S}_{\chi}$, is that they are always in a smooth
phase. This means that the symmetry of height translations is
always broken, and $\langle \chi_a \rangle = \langle h_a \rangle$
has some definite value across the entire system. As was argued
in Refs~\onlinecite{peierls1,peierls2}, any such definite value
necessarily breaks the lattice symmetry of the underlying
antiferromagnet, and will lead here to plaquette order.

With the assumption of a smooth interface, the optimal interface
configurations can be determine by a simple minimization of
${\cal S}_{\chi}$ by a set of time-independent values of $\chi_a$.
We allow for distinct expectation values, $\chi_W$, $\chi_X$,
$\chi_Y$, and $\chi_Z$ on the four dual sublattices. Then the
problem reduces to the minimization of the following energy as a
function of these four real variables:
\begin{eqnarray}
E_{\chi} &=& K \Bigl[ (\chi_X - \chi_W)^2 + (\chi_W-\chi_Y)^2
\nonumber
\\ &~&~~~~~+
(\chi_Y-\chi_Z)^2 + (\chi_Z-\chi_X)^2 \Bigr] \nonumber \\
&~&~~~~~- y_1 \left[ \cos(2 \pi \chi_W) -\cos(2 \pi \chi_Y)
\right] \nonumber \\ &~&~~~~~- y_2 \left[ \sin(2 \pi \chi_X) -
\sin(2 \pi\chi_Z) \right]
\end{eqnarray}
This minimization is a straightforward, but somewhat tedious,
computation. The present analysis is valid only for small $y_1$,
$y_2$, and so we analytically determine the minima in power
series in $y_{1,2}$. We define
\begin{eqnarray}
\chi_W &=& \chi_1 + \chi_2 + \chi_3 \nonumber \\
\chi_X &=& \chi_1 - \chi_2 + \chi_3 \nonumber \\
\chi_Y &=& \chi_1 + \chi_2 - \chi_3 \nonumber \\
\chi_Z &=& \chi_1 - \chi_2 - \chi_3.
\end{eqnarray}
We find that at the saddle points of $E_{\chi}$,
\begin{eqnarray}
\chi_2 &=& \frac{\pi^3 (y_1^2 + y_2^2)}{16 K^2} \sin(4 \pi \chi_1)
+ {\cal O} (y_{1,2}^4 ) \nonumber \\
\chi_3 &=& -\frac{\pi y_1}{2 K} \sin(2 \pi \chi_1) + {\cal O} (y_{1,2}^3) \nonumber \\
\chi_4 &=& \frac{\pi y_2}{2 K} \cos(2 \pi \chi_1) + {\cal O}
(y_{1,2}^3). \label{chis}
\end{eqnarray}
The average interface height, $\chi_1$, is determined by the
minimization of
\begin{equation}
E_{\chi} = E_0 + A \cos(4 \pi \chi_1) + B \cos( 8 \pi \chi_1) +
\ldots, \label{cosines}
\end{equation}
where $E_0$ is an uninteresting constant independent of $\chi_1$,
\begin{eqnarray}
A &=& \frac{\pi^2 (y_1^2 - y_2^2)}{2 K} - \frac{\pi^6 ( y_1^4 -
y_2^4)}{6 K^3} \nonumber \\
B &=& \frac{\pi^6( 7 y_1^4 + 6 y_1^2 y_2^2 + 7 y_2^4 )}{96 K^3},
\label{ab}
\end{eqnarray}
and all omitted terms are of order $y_{1,2}^6$ or higher (in
obtaining the results in (\ref{ab}) we had to include terms in
(\ref{chis}) which are one order higher than those shown). Note
that the square lattice antiferromagnet, with $y_1 = y_2$, has
$A=0$.

We now have to minimize (\ref{cosines}) to determine $\chi_1$.
Then from (\ref{chis}) we know $\chi_{2,3,4}$, and hence the
configuration of the interface heights. Then, from the connection
between $|h_a-h_b|$ and the corresponding dimer occupation
numbers, we can determine the pattern of the distribution
probabilities of the spin singlet bonds in the original
antiferromagnet. It is a simple exercise to determine the minima
of (\ref{cosines}) for different values of $A$ and $B$; the
resulting phase diagram is shown in Fig~\ref{figplaq}, and we now
list the various minima and the associated ground states of
the antiferromagnet.\\
({\em i\/}) $A\geq 0$, $B \leq A/4$: There are degenerate minima
at $\chi_1 = 1/4, 3/4$. The system spontaneously breaks a
translational symmetry by choosing one of these minima. With the
mappings above, it is easy to see that these are the plaquette
states, one of which is depicted in
Fig~\ref{figplaq}.\\
({\em ii\/}) $A\leq 0$, $B \leq -A/4$: Now the two equivalent
minima are $\chi_1 = 0, 1/2$. These also correspond to plaquette
states as above, but the chosen plaquettes are now around half of
those containing
diagonal links (see Fig~\ref{figplaq}).\\
({\em iii}) The remaining values of $A$ and $B$ have four
degenerate minima at $\chi_1 = 1/4 \pm \vartheta, 3/4 \pm
\vartheta$, where $0 < \vartheta < 1/4$ varies continuously as a
function of $A/B$. These states have spin-Peierls order of the
type shown in Fig~\ref{figplaq}: the links are divided into four
columnar sets, with each set having a different value of $\langle
{\bf S}_i \cdot {\bf S}_j \rangle$ on its links. This state
interpolates between the plaquette state in ({\em i\/}) as
$\vartheta\rightarrow 0$ and that in ({\em ii\/}) as $\vartheta
\rightarrow 1/4$.

The present analysis is for small $y_1$, and so, from (\ref{ab})
we should assume that $B \ll |A|$. Furthermore, the presence of
the frustrating $J_2$ interaction on half the plaquettes means
that the hedgehog tunneling events are more likely to be centered
on these plaquettes. Using the mapping of such events to the
model (\ref{schi}), we expect that $y_1 > y_2$. From (\ref{ab})
we therefore conclude that the most likely possibility for the
ground state is that in ({\em i\/}) above. The same state has
also been considered in Ref~\onlinecite{kk}.

We conclude this section with a few comments on the $(\pi, \pi)$
SRO phase of the antiferromagnet with full square lattice
symmetry, in which there is a diagonal $J_2$ exchange between
every pair of next-nearest-neighbor sites. Recent numerical work
on such an antiferromagnet\cite{wim,sushkov} has found evidence
for spin-Peierls ordering with the same spatial structure as in
({\em iii\/}) above for the Shastry-Sutherland antiferromagnet.
However, we noted earlier that the square lattice symmetry implies
that $A=0$: for this value, $\vartheta=1/8$, and the spin-Peierls
state of ({\em iii\/}) has a larger symmetry (two of the four sets
of columnar links are equal to each other), and becomes equivalent
to the ordering discussed in Refs~\onlinecite{peierls1,peierls2}.
To obtain $\vartheta \neq 1/8$, and so a ground state with the
symmetry of that in Fig~\ref{figplaq}, we need to add to
$E_{\chi}$ a higher order term $ C \cos(16 \pi \chi)$: then there
can be an {\em eight}-fold degenerate ground state, with
$\vartheta$ and $1/4-\vartheta$ equivalent to each other. This is
the state that appears to have been found in
Refs~\onlinecite{wim,sushkov}.

Note also that for the square lattice case, the $B<0$, $A=0$
solution has the four plaquette states degenerate with each
other\cite{cavo}.

\section{Deconfinement transition of the dimer phase}
\label{z2} The deconfined, ``spin-liquid'', $(\pi, q)$ SRO phase
in Fig~\ref{phase} is flanked on both sides by confining
paramagnetic phases, the plaquette and the dimer phases.

As we indicated Section~\ref{intro}, the deconfinement-confinement
quantum phase transition from the $(\pi, q)$ SRO phase to the
plaquette phase can be described in a theory essentially identical
to that considered previously for frustrated square lattice
antiferromagnets\cite{rsprl,rsijmpb}. At the mean-field level,
the transition is signaled by the onset of non-zero expectation
values of $Q_{ij}$ on the diagonal links: we will denote these
diagonal $Q_{ij}$ as $Q^d_{ij}$. Upon considering fluctuations, we
find that the $Q^d_{ij}$ constitute a charge 2 Higgs field in a
compact $U(1)$ gauge theory, and the deconfinment-confinement
transition is that in a $Z_2$ gauge
theory\cite{eduardo,rsprl,rodolfo,rsijmpb,wen,vojta,senthil}.

This section will consider the second deconfinement-confinement
transition in Fig~\ref{phase}, between the dimer and $(\pi, q)$
SRO phases in more detail. We will see that this is also described
by a $Z_2$ gauge theory, and the emergence of the $Z_2$ gauge
symmetry can be described in a somewhat more transparent manner.

As noted in Section~\ref{dimer}, the dimer phase is characterized
by non-zero expectation values of the diagonal $Q_{ij}^d$ links.
These links are all decoupled from each other, and this leads to a
simple, local structure in the effective action for the
fluctuations. The transition to the deconfined phase is now
signaled by the onset of non-zero expectation values of the
$Q_{ij}$ on the horizontal and vertical links, and we will denote
these by $Q_{ij}^h$ and $Q_{ij}^v$ respectively. Near the phase
boundary, we need only consider the structure of the effective
action as a functional of the $Q_{ij}^{h,v}$ after all other
degrees of freedom have been integrated out.

The simplest terms in the effective action arise from the on-site
propagation of the $b_i^{\alpha}$ on the site $i$ in imaginary
time. Integrating out the $b_{i}^{\alpha}$ in powers of the
$Q_{ij}^{h,v}$, the lowest order terms have the form
\begin{eqnarray}
{\cal S}_1 &=& \int d \tau \Bigl[ c_1 \sum_{\langle ij \rangle}
|Q_{ij}^{h,v}|^2
\nonumber \\
&+& c_2 \sum_{\Box} \left\{ Q_{12}^h Q_{23}^{v\ast} Q_{34}^h
Q_{41}^{v \ast} + \mbox{H.c.}\right\} + \ldots \Bigr] \label{s1}
\end{eqnarray}
where $c_1$, $c_2$ are constants, the first sum is over nearest
neighbor links, and the second sum is over plaquettes, with the
sites labeled as in Fig~\ref{figbox}. A crucial property of
${\cal S}_1$ is that all terms are invariant under a local $U(1)$
gauge transformation
\begin{equation}
Q_{ij}^{h,v} \rightarrow Q_{ij}^{h,v} e^{i (\phi_i + \phi_j)},
\label{u1gauge}
\end{equation}
where the phase $\phi_i$ can take arbitrary distinct values on the
sites $i$.

We have so far not made use of the fact that the nonzero value of
$\langle Q_{ij}^d \rangle$ allows the $b_{i}^{\alpha}$ bosons to
hop across a single diagonal link. Such hopping processes will
induce a large number of additional terms between the
$Q_{ij}^{h,v}$. We will now write down the structure of all such
terms which appear at fourth order in the $Q_{ij}^{h,v}$. It is
convenient to group these terms into sets associated with links
emanating from a given plaquette which does not have a diagonal
dimer across it: one such plaquette is that with the sites
1,2,3,4 in Fig~\ref{figbox}, and we now write down all four-link
terms in which every link has at least one site on the central
plaquette. It is not difficult to see that all other four-link
terms can be obtained by a simple translation of these terms to
other plaquettes. The terms are
\begin{eqnarray}
{\cal S}_2 = \int && d \tau \Biggl[ c_3 \Bigl\{ Q_{51}^{h}
Q_{26}^{v} Q_{32}^{v \ast} Q_{43}^{h} - Q_{26}^{v} Q_{37}^{h}
Q_{43}^{h \ast} Q_{41}^v  \nonumber \\ && + Q_{37}^{h} Q_{84}^{v}
Q_{41}^{v \ast} Q_{12}^h - Q_{84}^v Q_{51}^h Q_{12}^{h \ast}
Q_{32}^v + c.c. \Bigr\} \nonumber \\
&& + c_4 \Bigl\{ Q_{51}^{h} Q_{26}^{v} Q_{12}^{h \ast} Q_{41}^{v}
- Q_{26}^{v} Q_{37}^{h} Q_{32}^{v \ast} Q_{12}^h  \nonumber \\ &&
+ Q_{37}^{h} Q_{84}^{v} Q_{43}^{h \ast} Q_{32}^v - Q_{84}^v
Q_{51}^h Q_{41}^{v \ast} Q_{43}^h + c.c. \Bigr\} \nonumber \\ && -
c_5 \Bigl\{Q_{51}^h Q_{12}^{h \ast} Q_{37}^{h \ast} Q_{43}^h +
Q_{26}^{v} Q_{32}^{v \ast} Q_{84}^{v \ast} Q_{41}^v + c.c. \Bigr\}
\nonumber \\ && + c_6 \Bigl\{Q_{51}^h Q_{26}^v Q_{37}^h Q_{84}^v +
c.c.\Bigr\}\Biggr]. \label{s2}
\end{eqnarray}
Clearly, (\ref{s2}) is not invariant under (\ref{u1gauge}).
However, a residual $Z_2$ gauge symmetry does survive. We see that
(\ref{s1}, \ref{s2}), and all other allowed terms, are invariant
under
\begin{equation}
Q_{ij}^{h,v} \rightarrow Q_{ij}^{h,v} \eta_i \eta_j
\label{z2gauge}
\end{equation}
where $\eta_i = \pm 1$ performs the gauge transformation. However,
it is {\em not} possible to choose the $\eta_i$ independently on
every site: it is easy to see that we need the additional
constraint
\begin{eqnarray}
&& \mbox{$\eta_i=\eta_j$ whenever $i$ and $j$} \nonumber \\
&&~~~~\mbox{are separated by a diagonal link}. \label{z2cons}
\end{eqnarray}
So the $Z_2$ gauge degree of freedom is halved from that present
on the original square lattice.

To place the $Z_2$ gauge theory in a more conventional form, it is
useful to introduce a slightly different parameterization of the
degrees of freedom. First, we neglect all amplitude and phase
fluctuations and replace all the $Q_{ij}$ by discrete Ising
variables taken only the values $\pm 1$. Then we choose to
represent all the $Q_{ij}^{h}$ as Ising gauge fields, $\sigma$,
while all the $Q_{ij}^{v}$ are written as products of $\sigma$ and
a second Ising spin field, $\mu$; thus:
\begin{eqnarray}
Q^h & \sim & \sigma \nonumber \\
Q^v & \sim & \sigma \mu. \label{qz2}
\end{eqnarray}
This is shown a more explicitly in Fig~\ref{figbox}. Notice that
each pair of horizontal and vertical links that form a triangle
with a single diagonal link share the same Ising gauge field
$\sigma$. This choice is a consequence of the constraint
(\ref{z2cons})---as a result, all the $\mu$ fields are {\em
invariant} under the gauge transformation generated by the
$\eta_i$, while the $\sigma$'s transform like conventional Ising
gauge fields. This is also evident from the structure of the
effective action obtained by substituting the parameterization in
(\ref{qz2}) and Fig~\ref{figbox} into the effective action in
(\ref{s1}, \ref{s2}); for the terms displayed in (\ref{s1},
\ref{s2}) we obtain:
\begin{eqnarray}
{\cal S}_3 = \int  d \tau \Biggl[ && \widetilde{c}_2 \sigma_1
\sigma_2 \sigma_3 \sigma_4 \mu_2 \mu_4 \nonumber \\ &&+
\widetilde{c}_3 \sigma_1 \sigma_2 \sigma_3 \sigma_4 \Bigl\{ \mu_1
\mu_2 - \mu_1
\mu_4  + \mu_3 \mu_4 - \mu_3 \mu_2 \Bigr\} \nonumber \\
&& + \widetilde{c}_4 \Bigl\{ \mu_1 \mu_4 - \mu_1 \mu_2 + \mu_2
\mu_3 - \mu_3 \mu_4 \Bigr\} \nonumber \\ && - \widetilde{c}_5
\sigma_1 \sigma_2 \sigma_3 \sigma_4 \Bigl\{1 + \mu_1 \mu_2 \mu_3
\mu_4 \Bigr\} \nonumber \\ && + \widetilde{c}_6 \sigma_1 \sigma_2
\sigma_3 \sigma_4 \mu_1 \mu_3 \Biggr]. \label{sfinal}
\end{eqnarray}
The terms involving the $\sigma_i$ appear to have the plaquette
form associated with Ising gauge fields. The spatial structure of
these gauge interactions is made clearer by the transformation in
Fig~\ref{figz2}. Here, we have collapsed pairs of sites connected
by the diagonal links into single sites---we now see that the
$\sigma_i$ can viewed as residing on the links of a square lattice
which is tilted by $45^{\circ}$ from the original lattice, and
their gauge interactions have the usual form around elementary
plaquettes.

The $\mu_i$ constitute a separate global Ising degree of freedom
associated with the breaking of the symmetry of $90^{\circ}$
spatial rotations between the horizontal and vertical directions.
In the mean-field theory of the deconfined phase, the state with
$\mu_i = 1$ corresponds to the state with dominant spin
correlations at the wavevector $(\pi, q)$ (say). The degenerate
partner state with spin correlations at $(q, \pi)$ is obtained by
the state $\mu_i = (-1)^{i_y}$, where $(i_x, i_y)$ are the
Cartesian co-ordinates of the site $i$.

So the action ${\cal S}_3$ describes a $Z_2$ gauge theory
($\sigma$) coupled (rather intricately) to an Ising spin field
($\mu$); the $\mu$ field does not carry a non-zero charge under
the $Z_2$ gauge transformation. The $Z_2$ gauge theory can undergo
a confinement-deconfinement transition (which is related by a
duality transformation to the magnetic transition in an Ising
model in three dimensions), corresponding to the liberation of
spinons upon moving out of the dimer phase. In a different sector,
the ordering of the $\mu$ degrees of freedom leads to the
appearance of nematic order, and the breaking of the symmetry of
$90^{\circ}$ spatial rotations. In the mean-field theory, these
two transitions occur at the same point {\em i.e.} the
deconfinement transition is also the point where the spatial
rotation symmetry is broken. More generally, the interplay between
these two potentially distinct transitions can be addressed by an
analysis of fluctuations using the action ${\cal S}_3$. It does
appear possible that the two transitions are not simultaneous, and
that there can be a deconfined phase without any broken spatial
symmetries; moreover, if there is a simultaneous transition in the
two sectors, it is likely to be first order. A more definitive
conclusion on these issues must await a complete study of the
coupled Ising gauge/Ising spin theory defined by ${\cal S}_3$. We
note that these issues concerning the transition from the
confined dimer phase to the deconfined  helical SRO phase are
somewhat different from earlier deconfinement
transitions\cite{vojta} because here the dimer phase does not
break any lattice symmetries.

\section{Conclusions}
\label{conc}

The Mott insulator SrCu$_2$(BO$_3$)$_2$ is perhaps the only
example of a spin gap paramagnet on a strongly frustrated
two-dimensional lattice (another example of a two-dimensional
paramagnet is CaV$_4$O$_9$, but its spin gap is realized by
dilution and not frustration). To date, it appears that the spin
gap is realized in a simple decoupled dimer ground state
discovered originally by Shastry and Sutherland\cite{Shastry}.
Here, we undertook a more detailed study of the parameter space of
this antiferromagnet, and found that other paramagnetic spin gap
states are also possible. One of these was the plaquette
state\cite{kk}, which appears in a region of weaker frustration
and commensurate spin correlations. The other was a more exotic
state with ``topological order'', deconfined $S=1/2$ excitations,
and helical spin correlations. The latter state was found to be
contiguous to the dimer state, and so not too far from the
physically relevant regime: it appears that SrCu$_2$(BO$_3$)$_2$
is quite close to the boundary of stability of the dimer phase.

Our results suggest exciting possibilities for materials obtained
by doping SrCu$_2$(BO$_3$)$_2$ with mobile carriers. It is
expected that the helical state will be more amenable to the
motion of charge carriers than the dimer state, and so doping may
well drive the system into a topologically ordered state. Such a
state is a prime candidate for superconductivity with the exotic
properties associated with the proximity of a Mott insulator with
deconfined spinons: these include the flux-trapping effect of
Senthil and Fisher\cite{senthil2}, and a regime of stable $hc/e$
vortices\cite{ss,nl}. An experimental effort to dope
SrCu$_2$(BO$_3$)$_2$ (or related compounds) therefore appears
worthwhile.


\acknowledgements

This research was supported by US NSF Grant Nos. DMR 96--23181 (SS) and
DMR 97--12391 (JBM).


\appendix

\section*{}

\begin{figure}
\vspace{1in}
\epsfxsize=3in \centerline{\epsffile{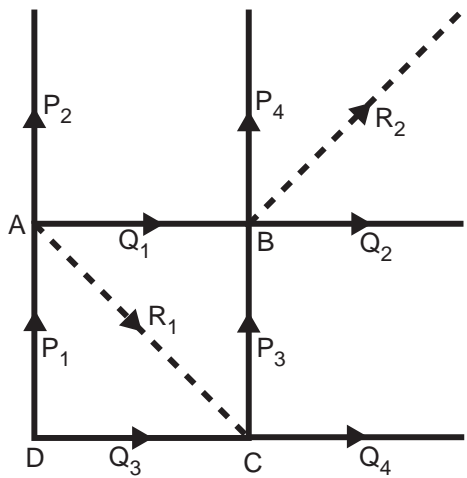}}
\caption{The four sites of the unit cell (labeled A, B, C and D), and
the 10 link variables $Q_{ij}$.}
\label{unit}
\end{figure}

\begin{figure}
\vspace{1in} \epsfxsize=3.2in \centerline{\epsffile{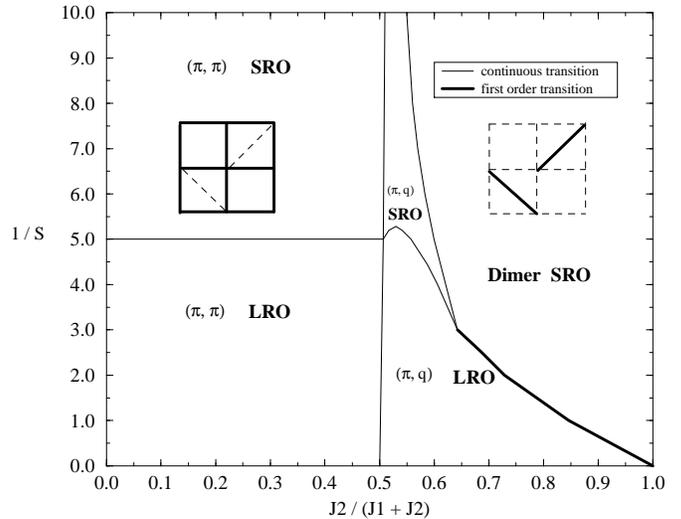}}
\caption{Large $N$ phase diagram of the Sp(N) Shastry-Sutherland
model, Eq. \ref{SSH}, as a function of $J_2/(J_1+J_2)$ and $1/S$.
The five phases are described in Section~\protect\ref{mft}. The
LRO phases break spin-rotation symmetry: the spin order is
collinear and commensurate in the $(\pi,\pi)$ LRO phase, and
helical and incommensurate in the $(\pi, q)$ LRO phase. The SRO
phases preserve spin rotation invariance. In the $(\pi,\pi)$ SRO
only the horizontal and vertical $Q_{ij}$ are non-zero in the
large $N$ theory--fluctuations lead to broken translational
symmetry in one of the states shown in Fig~\protect\ref{figplaq}.
(A state with co-existing $(\pi,\pi)$ LRO and plaquette order is
also allowed by the theory \protect\cite{rodolfo2} beyond the
large $N$ limit (not shown above), and there is evidence that this
occurs on a frustrated square lattice antiferromagnet
\protect\cite{sushkov}.) The dimer phase has only the diagonal
$Q_{ij}$ non-zero in the large $N$ theory. The $(\pi, q)$ SRO
phase has all the $Q_{ij}$ non-zero: this phase has topological
order and deconfined spinons. } \label{phase}
\end{figure}

\begin{figure}
\epsfxsize=3.1in \centerline{\epsffile{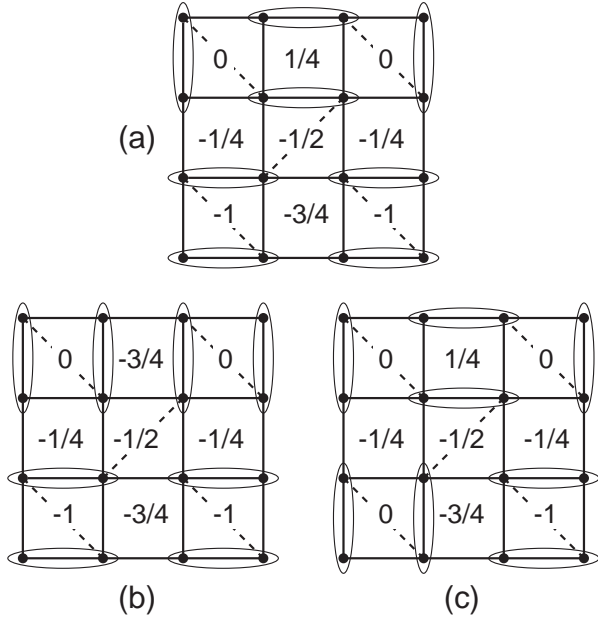}}
\caption{Three states of the Hilbert space of the quantum dimer
model. There are off-diagonal matrix elements in the effective
Hamiltonian which connect state (a) to state (b), and state (a)
to state (c), by a resonance between pairs of horizontal and
vertical dimers around a plaquette. The latter matrix element
differs from the former because only the latter has a diagonal
link across the resonating plaquette. Also shown are the
corresponding values of the heights, $h_a$, on the sites of the
dual lattice.} \label{figdimer}
\end{figure}

\begin{figure}
\epsfxsize=1.6in \centerline{\epsffile{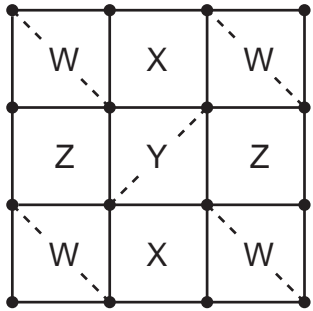}}
\vspace{12pt} \caption{The four dual sublattices upon which the
height offsets take the values $\zeta_W = 0$, $\zeta_X=1/4$,
$\zeta_Y = 1/2$, and $\zeta_Z = 3/4$.} \label{figzeta}
\end{figure}

\begin{figure}
\epsfxsize=3.5in \centerline{\epsffile{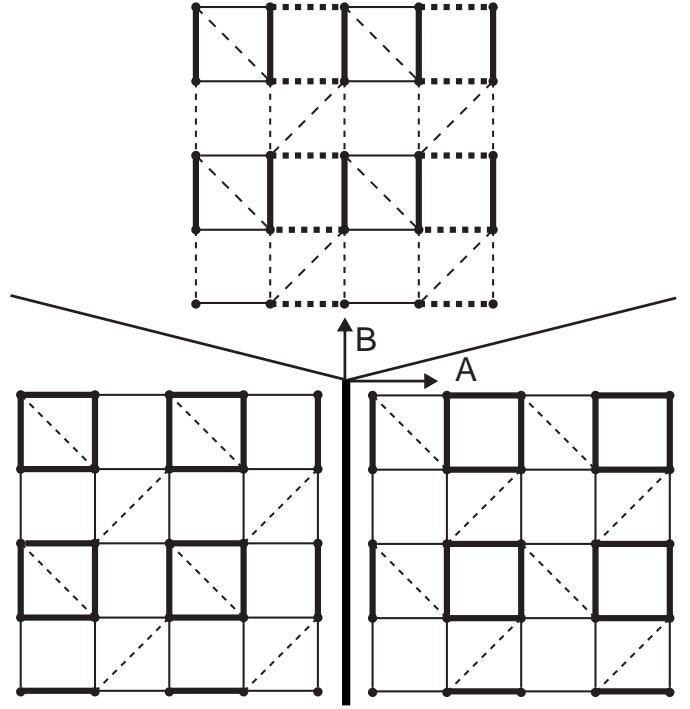}} \vspace{12pt}
\caption{Phase diagram of (\protect\ref{cosines}) as a function of
the parameters $A$ and $B$; this model describes fluctuations in
the $(\pi,\pi)$ SRO phase of Fig~\protect\ref{phase}. The thick
line is a first order transition, while the thin lines are second
order. The plaquette and spin-Peierls states are shown, with the
different line-styles representing distinct values of $\langle
{\bf S}_i \cdot {\bf S}_j \rangle$ across the links.}
\label{figplaq}
\end{figure}

\begin{figure}
\epsfxsize=2.9in \centerline{\epsffile{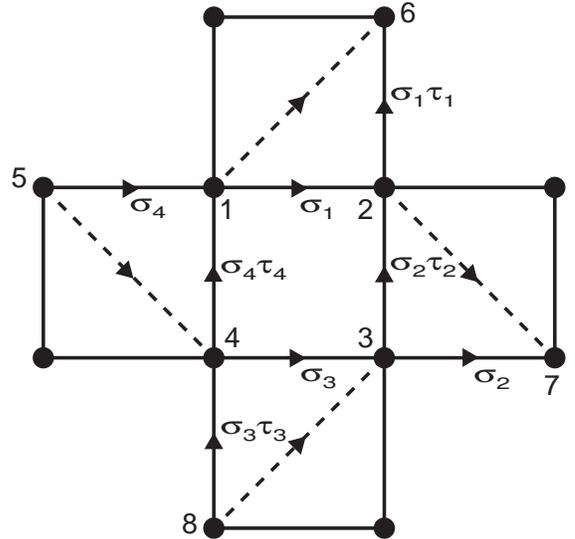}} \caption{ A
section of the Shastry-Sutherland lattice. We have labeled sites
around the central plaquette to enable the discussion in
Section~\ref{z2} of the various terms in the $Z_2$ gauge theory of
the transition from the dimer state to the $(\pi, q)$ SRO phase
with spinon deconfinement.} \label{figbox}
\end{figure}

\begin{figure}
\epsfxsize=2.7in \centerline{\epsffile{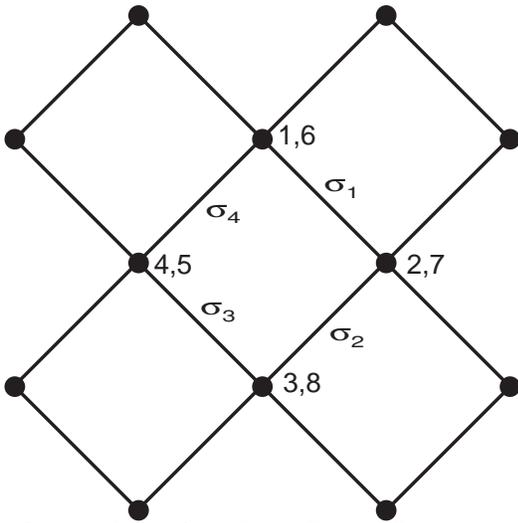}} \caption{A
deformation of the Shastry-Sutherland lattice which exposes the
structure of the $Z_2$ gauge theory. Pairs of sites across a
diagonal bond have been compressed into a single site. Four of the
sites carry pairs of sites labels, corresponding to the original
site numbers in Fig~\protect\ref{figbox}. The $Z_2$ Ising gauge
fields on some of the links are indicated, with a notation
corresponding to the degrees of freedom in
Fig~\protect\ref{figbox}.} \label{figz2}
\end{figure}

This appendix will provide a proof of a statement made in
Section~\ref{helical} on the nature of the saddle point in the
phase with helical LRO: we will analytically show that in the
$(\pi, q)$ and $(q, \pi)$ LRO phase the link fields obey the
following relations:
\begin{eqnarray}
Q &=& P \nonumber \\
|R_1| &=& |R_2|,
\label{PQR}
\end{eqnarray}
where $|Q_i| = Q$, $|P_i| = P$. The reasoning is the same in the
both $(\pi, q)$ and $(q, \pi)$ phase, and we will fix our state in
the $(\pi, q)$ phase for simplicity. In this state, the directions
of the link fields $Q_{ij}$ are shown as in Fig~\ref{unit}. The
spinon dispersion in this phase can be obtained from the following
eigenvalue equation\cite{triangle} in momentum space:
\begin{eqnarray}
\tau^3 D(k) M &=& M\tau^3\hat{\omega}(k)\nonumber \\
\tau^3 &=&
\left(\begin{array}{cc}
\P & 0 \\
0 & -\P
\end{array}\right)\nonumber \\
D(k) &=&
\left(\begin{array}{cc}
\lambda\P& P(k) \\
P^{\dagger}(k) & \lambda\P
\end{array} \right)
\label{eigeneqn}
\end{eqnarray}
where $\P$ is the $4 \times 4$ unit matrix, $\hat{\omega}(k)$ is a
diagonal matrix of the bosonic eigenenergies, $M$ is a $8\times 8$
matrix whose columns are the eigenvectors of the matrix $\tau^3 D(k)$
and the diagonal elements of $\tau^3\hat{\omega}(k)$ are the corresponding
eigenvalues, $P(k)$ is a $4 \times 4$ matrix with the following form
\begin{eqnarray}
P(k) &=& \left( \begin{array}{cc} 0  &{\it{i}}J_1Q\sin(k_x) \\
{\it{i}}J_1Q\sin(k_x) & 0   \\
-(J_2 R_1 /2)e^{-{\it{i}}(k_x-k_y)}&{\it{i}}J_1P\sin(k_y) \\
{\it{i}}J_1P\sin(k_y) &-(J_2 R_2 /2)e^{-{\it{i}}(k_x+k_y)}
\end{array}\right. \nonumber \\
&~&~~~~~~~~~~\left. \begin{array}{cc} (J_2 R_1 /
2)e^{{\it{i}}(k_x-k_y)}&
{\it{i}}J_1P\sin(k_y) \\
(J_2 R_2 /2)e^{{\it{i}}(k_x+k_y)} \\ 0 &
{\it{i}}J_1Q\sin(k_x) \\
{\it{i}}J_1Q\sin(k_x) &0
\end{array}\right),
\label{P(k)}
\end{eqnarray}
and $\lambda$ is the Largrange multiplier of the mean-field
Hamiltonian which we assume to be independent of lattice site
$i$. With this assumption, it can be shown\cite{triangle} that the
eigenvalues occur in pairs with opposite signs
$(\omega_{\mu}(k), -\omega_{\mu}(k))$ where $\mu = 1,\cdots,4$, and
the matrix $M$ has the form
\begin{equation}
M = \left( \begin{array}{cc}
U & -V^{\ast}\\
V & U^{\ast}
\end{array}\right),
\label{M}
\end{equation}
where the $U$, $V$ are $4 \times 4$ matrices associated with the
positive eigenvalues. The $(\pi, q)$ LRO phase ($x_{i}^{\alpha}\not=0$)
occurs at the wavevector $\vec{k}_{min} = (\pm\pi/2, \pm q/2)$ where
the eigenenergy vanishes, ie. $\omega(\vec{k}_{min}) = 0$.
The two linearly independent eigenvectors associated with
$\vec{k}_1 = (\pi/2, q/2)$ and $\vec{k}_2 = (\pi/2, -q/2)$ can be
shown to be
\begin{eqnarray}
\Psi_1 &=& (1, {\it{i}} e^{-{\it{i}}q/2}, {\it{i}} e^{-{\it{i}}q/2}, 1,
{\it{i}}, -e^{-{\it{i}}q/2}, -e^{-{\it{i}}q/2}, {\it{i}})
e^{{\it{i}}\vec{k}_1\cdot\vec{r}}
\nonumber \\
\Psi_2 &=& (1, {\it{i}} e^{{\it{i}}q/2}, {\it{i}} e^{{\it{i}}q/2}, 1,
-{\it{i}}, e^{{\it{i}}q/2}, e^{{\it{i}}q/2}, -{\it{i}})
e^{{\it{i}}\vec{k}_2\cdot\vec{r}}
\label{eigenvec}
\end{eqnarray}
respectively. Substituting $\Psi_1$ (or $\Psi_2$) into (\ref{eigeneqn}),
we have
\begin{eqnarray}
\lambda &-& [J_1 (P + Q) \sin(q/2) + J_2 \frac{R_1}{2} \sin(q)]
\nonumber \\ &+&
         {\it{i}}[J_1 Q \cos(q/2) + J_2 \frac{R_1}{2} \cos(q)] = 0\nonumber \\
\lambda &-& [J_1 (P + Q) \sin(q/2) + J_2 \frac{R_2}{2} \sin(q)]
\nonumber \\ &+&
         {\it{i}}[J_1 Q \cos(q/2) + J_2 \frac{R_2}{2} \cos(q)] = 0
\label{R1R2}
\end{eqnarray}
We an easily see that $R_1 = R_2 = R$ from the above conditions.
Also, we find that each saddle-point may be described by purely real
$<Q_{ij}>$. Therefore, we may fix the values of $\lambda$ and $q$ in
the LRO phase from the above condition.
\begin{eqnarray}
J_1 (P + Q) \sin(q/2) + J_2 \frac{R}{2} \sin(q) = \lambda  \nonumber \\
J_1 Q \cos(q/2) + J_2 \frac{R}{2} cos(q)  =  0
\label{lambdaq}
\end{eqnarray}
To prove $P = Q$, we need one additional condition from the
saddle-point equations. The mean-field free energy $E_{MF}$
is a function of $\lambda$, $Q$, $P$, $R$ and $x^{\alpha}(q)$ where
these are independent parameters. The large-N solutions of this
model are obtained by solving the saddle-point equations which
set the derivatives of free energy with respect to these independent
variables to be zero. Notice that $q$ is also an independent
parameter. The additional condition we need comes from the
saddle-point equation associated with $q$. It is given by
\begin{equation}
\frac{\partial{E_{MF}}}{\partial{q}} = 0.
\label{MFEq}
\end{equation}
The only explicit $q$ dependence in the free energy is in
the bose condensate variables $x^{\alpha}(q)$. This piece of
free energy is given by\cite{triangle}
\begin{equation}
E_{x}(q) = \sum_{i>j}\frac{J_{ij}}{2}\left[-Q_{ij}{\epsilon}_
{\sigma\sigma^{\prime}} x_{i}^{\sigma} (q) x_{j}^{\sigma^{\prime}}(q)
 + \mbox{H.c.}\right],
\label{Exq}
\end{equation}
where ${\epsilon}_{\sigma\sigma^{\prime}}$ is the $SU(2)$ antisymmetric
$\epsilon$ tensor, and $\sigma,\sigma^{\prime}=\uparrow,\downarrow$.
The condensates $x_i^{\sigma}(q)$ must be the linear combinations of
the eigenvectors $\Psi_1$ and $\Psi_2$ associated with the zero mode:
this introduces two complex numbers $c_1$, $c_2$, with only the value
of $|c_1|^2 + |c_2|^2$ fixed by the saddle-point equations\cite{triangle}.
Working out the orientation of the condensate at every lattice
site over the unit cell, the condensates can be written in the form
\begin{eqnarray}
\left(\begin{array}{c}
x_{A}^{\uparrow}\\
x_{A}^{\downarrow}
\end{array} \right) &=&
\left(\begin{array}{c}
c_1 + c_2\\
{\it{i}}c_2^{\ast} - {\it{i}}c_1^{\ast}
\end{array}\right) \nonumber \\
\left(\begin{array}{c}
x_{B}^{\uparrow}\\
x_{B}^{\downarrow}
\end{array} \right) &=&
\left(\begin{array}{c}
-c_1e^{-{\it{i}}q/2} - c_2e^{{\it{i}}q/2}\\
-\it{i}c_2^{\ast}e^{-{\it{i}}q/2} + {\it{i}}c_1^{\ast}e^{{\it{i}}q/2}
\end{array}\right)\nonumber \\
\left(\begin{array}{c}
x_{C}^{\uparrow}\\
x_{C}^{\downarrow}
\end{array} \right) &=&
\left(\begin{array}{c}
-c_1e^{-{\it{i}}q} - c_2e^{{\it{i}}q}\\
-{\it{i}}c_2^{\ast}e^{-{\it{i}}q} - {\it{i}}c_1^{\ast}e^{{\it{i}}q}
\end{array}\right)\nonumber \\
\left(\begin{array}{c}
x_{D}^{\uparrow}\\
x_{D}^{\downarrow}
\end{array} \right) &=&
\left(\begin{array}{c}
c_1e^{-{\it{i}}q/2} + c_2e^{{\it{i}}q/2}\\
{\it{i}}c_2^{\ast}e^{-{\it{i}}q/2} - {\it{i}}c_1^{\ast}e^{{\it{i}}q/2}
\end{array}\right)
\label{xABCD}
\end{eqnarray}
By substituting (\ref{xABCD}) into (\ref{Exq}), we can explicitly work
out $E_x(q)$ . It is given by
\begin{equation}
E_x(q) = -[J_1 (P + Q)\sin(q/2) + J_2 \frac{R}{2}\sin(q)] (|c_1|^2+|c_2|^2).
\label{finalExq}
\end{equation}
Now the saddle-point condition (\ref{MFEq}) becomes
\begin{equation}
\frac{\partial{E_{MF}}}{\partial{q}} =
J_1 \frac{P + Q}{2}\cos(q/2) + J_2 \frac{R}{2}\cos(q) = 0.
\label{finalMFEq}
\end{equation}
Combining (\ref{lambdaq}) and (\ref{finalMFEq}), we have $P = Q$.



\end{document}